\documentclass[prd,preprint,tightenlines,floatfix,showpacs,preprintnumbers,nofootinbib,eqsecnum]{revtex4-1}

\pdfoutput=1

\usepackage[dvips,final]{graphicx}
\usepackage{color,xcolor}
\usepackage{amssymb}
\usepackage{amsmath}
\usepackage{amsfonts}
\usepackage{epsfig}
\usepackage{bm}
\usepackage{comment}
\usepackage{float}
\usepackage[utf8]{inputenc}
\usepackage{comment}

\colorlet{RED}{red}

\def\qq{$q\bar{q}$ }

\DeclareSymbolFont{myletters}{OML}{ztmcm}{m}{it}
\DeclareMathSymbol{\uplambda}{\mathord}{myletters}{"15}

\begin{document}

\title{Exclusive photo- and electroproduction of excited light vector 
mesons via holographic model}

\author{Cheryl Henkels$^{1}$}
\email{cherylhenkels@hotmail.com}

\author{Emmanuel G. de Oliveira$^{1}$}
\email{emmanuel.de.oliveira@ufsc.br}

\author{Roman Pasechnik$^{2}$}
\email{Roman.Pasechnik@thep.lu.se}

\author{Haimon Trebien$^{1}$}
\email{haimontrebien@outlook.com}

\affiliation{
\\
{$^1$\sl Departamento de F\'isica, CFM, Universidade Federal 
de Santa Catarina, C.P. 476, CEP 88.040-900, Florian\'opolis, 
SC, Brazil
}\\
{$^2$\sl
Department of Astronomy and Theoretical Physics, Lund
University, SE-223 62 Lund, Sweden
}}

\begin{abstract}
\vspace{0.5cm}
In this paper, we study total and differential observables of electro- and photoproduction 
of light $\rho$, $\omega$ and $\phi$ mesons as functions of the center-of-mass energy 
of the $\gamma p$ collision and momentum transfer squared $|t|$. The corresponding vector mesons 
wave functions have been computed in the framework of relativistic AdS/QCD holographic approach. 
A satisfactory description of all available data on ground-state $\rho(1S)$, $\omega(1S)$ and $\phi(1S)$ 
mesons production cross sections has been achieved in the color dipole picture. Finally, the key 
observables of excited $\rho(2S)$, $\omega(2S)$ and $\phi(2S)$ states production in $\gamma^{(*)} p$ collisions 
have been presented here using a common wave function formalism. This study reveals a large theoretical uncertainty coming from 
the modeling of the partial dipole amplitude in the nonperturbative kinematical domain. 
Hence, the latter could benefit from future measurements of photoproduction of the excited states.
\end{abstract}


\maketitle

%%%%%%%%%%%%%%%%%%%%%%%%
\section{Introduction}
\label{Sect:intro}
%%%%%%%%%%%%%%%%%%%%%%%%

Historically, an important milestone for particle physics was associated with thorough experimental studies of the proton internal structure by the HERA collider at DESY \cite{Glazov:2007zz,H1:2009pze,H1:2015ubc, Fagundes:2022bzw}. Such measurements have collected, along many years, an extensive amount of exclusive vector meson production data providing a very detailed picture of the proton through various differential observables carrying detailed information about its transverse shape \cite{Mantysaari:2020axf}. The corresponding processes have also served as an important tool to explore the perturbative limit of QCD at a hard scale given by the mass of heavy vector mesons \cite{Ivanov:2004ax,Brambilla:2010cs,Henkels:2020kju,Henkels:2020qvo} enabling an approximate separation of perturbative dynamics from non-perturbative effects known as factorisation. However, when the objects of study are light vector mesons, their mass is not high enough to provide a sufficiently hard scale for factorisation to work. In this case, only effective models can be applied for making estimates of the corresponding observables in the essentially soft regime of QCD \cite{Nemchik:1994fp,Nemchik:1996cw}.

In particular, the light vector meson production processes can be studied in the color dipole picture \cite{Kopeliovich:1981pz, SampaiodosSantos:2014qtt, Goncalves:2017wgg,Goncalves:2020cir,Nemchik:1996cw,Nemchik:2000dd}. It is based on the fact that in the proton rest frame the lifetime of a quark-antiquark fluctuation of the projectile photon is bigger than the proton radius. This permits to separate the process into three steps: first the incoming photon fluctuates into its leading Fock state, a color-neutral \qq pair known as the color dipole. Then, such a color dipole propagates through the color field of the proton target and interacts with it by means of an exchange of a gluonic ladder. The partial dipole amplitude of the dipole-target scattering carries the relevant information about all the non-perturbative effects associated with the soft gluons interactions inside the proton, and hence, about the proton structure \cite{Hufner:2000jb}. Such effects can be separated in a universal way such that the corresponding dipole cross section is often parameterized and fitted to the available data \cite{Kopeliovich:2001xj}. The observables obtained within this approach are very sensitive to the parameterization of the dipole cross section, especially, in kinematical domains that were not sufficiently well probed in existing (e.g. HERA) measurements. In this work, we explore exclusive light vector meson $\rho$, $\omega$ and $\phi$ production in the dipole picture employing two distinct models for the impact parameter dependent partial dipole amplitude that best describe the exclusive vector meson production data from the HERA collider as has been discussed earlier in Refs.~\cite{Henkels:2020kju, Henkels:2020qvo}. The last step is the projection of the \qq pair into a vector meson which is encoded in the vector meson wave function. 

There is a big challenge to obtain the light vector meson wave functions from first principles, especially for light vector mesons. For the case of light quarks, the relativistic and non-perturbative effects dominate in the vector meson wave function, and a rigorous theoretical framework consistently taking those into account is needed. In the current analysis, we work with the phenomenologically successful model of relativistic AdS/QCD holographic wave functions for light vector mesons extensively discussed in Refs.~\cite{Brodsky:2008pg,Brodsky:2014yha,Forshaw:2012im}. The AdS/QCD holographic wave function is based on the correspondence between string states in anti-de Sitter space (AdS) and conformal field theories (CFT) in Minkowski spacetime \cite{Brodsky:2008kp}. The generalization of this correspondence to QCD is non-trivial due to dynamical breaking of the conformal symmetry in QCD. However, there are certain empirical and theoretical arguments that support this generalization (see e.g.~Refs.~\cite{Csaki:2008dt,Vega:2008te}). A particular advantage of this model is that it enables to incorporate the quark confinement effects through an effective potential, and then obtain the realistic vector meson wave function by solving a relativistic equation in the AdS space. The only free parameter in this model $\kappa$ is related to the strength of the quarks confinement. 

One advantage of this approach is that the solutions for the relativistic light-front equation are frame independent \cite{Brodsky:2014yha, RevModPhys.21.392}. Since the amplitudes in the dipole model are also calculated using the light--front variables, the hadron wave functions obtained with the holographic approach are very convenient. Besides that, this approach has been widely used in many applications, such as in the determination of the Drell-Yan-West formula \cite{PhysRevLett.24.181, PhysRevLett.24.1206}, the description of hadron spectra \cite{Karch:2006pv,Brodsky:2006uqa}, phenomenological studies of QCD couplings \cite{Mattingly:1993ej,Brodsky:2002nb, Baldicchi:2002qm}, description of weak hadron decays \cite{Hambye:2005up}.

It is known to be very difficult to describe all existing data for $\rho$, $\omega$ and $\phi$ mesons simultaneously in a single framework \cite{Goncalves:2017wgg, Ahmady:2016ujw}, using a common dipole and wave function formalism. In this work, we found that, if $\kappa$ depends on the ground-state meson mass (i.e. when it is not assumed to be a universal parameter), one can reach a rather good description of all the available data on the corresponding production cross sections in the color dipole approach. Such a success of our analysis has permitted us to make predictions for the key observables of excited $\rho(2S)$, $\omega(2S)$ and $\phi(2S)$ states' photo- and electroproduction, whose discussions and predictions are still scarce in the literature. The existing and rather old estimates make use of gaussian vector meson wave functions with modifications for the excited states, see e.g.~Refs.~\cite{Nemchik:1996cw,Nemchik:2000dd}, and need to be revisited.

The article is organised as follows. In Sect.~\ref{sec:formalism}, we have given a short description of the differential and total cross section of elastic vector meson photoproduction $\gamma p \rightarrow V p $ off the proton target in terms of the holographic LF wave functions of the mesons and the dipole models employed in this study. Sect.~\ref{sec:results} presents the numerical calculations for the differential cross section of the $\gamma p \rightarrow V p$ process for the ground and excited states of $\rho$, $\omega$ and $\phi$ mesons, with results successfully describing the existing data for the ground states but still very sensitive to the dipole model used. At last, a brief summary of our results is given in Sect. \ref{sec:conclusions}.

%--------------------------------
\section{Theoretical formalism}
\label{sec:formalism}
%--------------------------------

%------------------------------------------------------------------------------------------
\subsection{Exclusive production amplitude of light vector mesons in $\gamma p$ collisions}
%------------------------------------------------------------------------------------------

Considering the proton target case, the exclusive diffractive differential cross section for the $\gamma p \rightarrow V p$ process of vector meson $V$ (with mass $M_V$) production reads:
\begin{equation}
    \frac{d \sigma^{\gamma p \rightarrow V p}}{d t} = \frac{1}{16 \pi } |\mathcal{A}^{\gamma p \rightarrow V p} (x, \Delta_T) |^2 \,,
    \label{amplitude_diff}
\end{equation}
where $t = - \Delta_T^2$ is the momentum transfer squared, $\Delta_T \equiv |\mathbf{\Delta}|$ is the transverse momentum of the produced vector meson recoiled against the proton target, and $\mathcal{A}^{\gamma p \rightarrow V p }$ is the elastic production amplitude given by
\begin{equation}
    \mathcal{A}^{\gamma p \rightarrow V p}(x,\Delta_T) = \int d^2 \mathbf{r} \int_0^1 d \beta (\Psi_V^* \Psi_{\gamma}) \mathcal{A}_{q \bar{q}} (x, \mathbf{r}, \mathbf{\Delta})\,.
    \label{amplitude}
\end{equation}
This amplitude is the product of the elastic elementary amplitude $\mathcal{A}_{q \bar{q}}$ with an overlap between the photon wave function ($\Psi_{\gamma}$) and the vector meson wave function ($\Psi_V$), integrated over the size of the dipole $\textbf{r}$ and the longitudinal momentum carried by the quark $\beta$. In this work we applied the purely perturbative photon wave function coming from QED \cite{Forshaw:2003ki}, commonly used in the literature, while attributing all non-perturbative QCD effects to the partial dipole amplitude. It is worth mentioning here previous works \cite{Goncalves:2020cir, Kopeliovich:1999am, Kopeliovich:2001xj} that modify the photon wave function for large dipole size to incorporate non-perturbative QCD effects, especially considering production of light vector mesons with low photon virtuality. The light vector meson wave function $\Psi_V$ will be discussed in a section below.

One can write straightforwardly the imaginary part of the elastic scattering amplitude in terms of the partial dipole amplitude $N(x, \mathbf{r}, \mathbf{b} ) \equiv \mathrm{Im} \mathcal{A}_{q \bar q} (x, \mathbf{r}, \mathbf{b} ) = 2 [1 - \mathrm{Re} S (x, \mathbf{r}, \mathbf{b})]\, $, such that the corresponding amplitude can be represented as follows \cite{Kowalski:2006hc}:
\begin{equation}
    \mathcal{A}^{\gamma p \rightarrow V p} (x, \Delta_T) = 2 i \int d^2 \mathbf{r} \int_0^1 d \beta \int d^2 \mathbf{b} (\Psi_V^* \Psi_{\gamma}) e^{-i[\mathbf{b} -(1-2 \beta) \mathbf{r}/2]\cdot \mathbf{\Delta}} N(x, \mathbf{r}, \mathbf{b} )\,.
    \label{hatta correction}
\end{equation}
Here, in the non-forward case, the phase $e^{-i(1-2 \beta)\frac{\mathbf{r}}{2}\cdot \mathbf{\Delta}}$ is slightly different from the one found in Ref.~\cite{Kowalski:2006hc} as it has been corrected in more recent studies to account for $\beta \rightarrow 1 - \beta$ symmetry between the quark and the antiquark momentum fractions \cite{Hatta_2017}. 

The total cross section of exclusive vector meson production in $\gamma p$ collisions,
\begin{equation}
    \sigma^{\gamma p \rightarrow V p} (W)= \frac{1}{16 \pi B}   |\mathcal{A}^{\gamma p \rightarrow V p} (W) |^2 \,,
\end{equation}
depends on the $\gamma p$ center-of-mass energy $W$ through $x = (Q^2 + M_V^2)/(Q^2 + W^2)$. In this work, we use the elastic slope parameter $B$ which is in accordance to the available data from the ZEUS Collaboration~\cite{Forshaw:2010py,ZEUS:2007iet}, such that the low-$t$ dependence of the electroproduction can then be consistently restored in the exponential form. It takes the form
\begin{equation}
    B = N \left[ 14.0 \left( \frac{1 \mathrm{GeV}^2}{Q^2 + M_V^2}\right)^{0.2} + 1 \right] \,,
\end{equation}
with $N = 0. 55 \, \mathrm{GeV}^{-2}$.

In order to take into account the real part of the $\mathcal{A}_{q \bar q}$ amplitude, it is necessary to introduce in Eq.~\eqref{amplitude_diff} a factor that represents the ratio of the real to imaginary parts of the exclusive production amplitudes as follows \cite{Hufner:2000jb}:
\begin{equation}
    \mathcal{A}^{\gamma p \rightarrow V p} \Rightarrow \mathcal{A}^{\gamma p \rightarrow V p} \left( 1 - i \frac{\pi \lambda}{2} \right)\,, \quad \mathrm{with} \quad \lambda = \frac{\partial \ln \mathcal{A}^{\gamma p \rightarrow V p}}{\partial \ln (1/x)}\,.
    \label{realpart}
\end{equation}

We also included the so-called skewness effect, which takes into account the fact that the gluons exchanged between the $q \bar{q}$ pair and the target can carry different momentum fractions from the target. This effect is included via a multiplicative factor $R^2_g$ applied to the differential cross section in Eq.~\eqref{amplitude_diff} given by (see e.g.~Ref.~\cite{Shuvaev:1999ce} for further details)
\begin{equation}
    R_g^2 (\lambda) = \frac{2^{2 \lambda + 3}}{\sqrt{\pi}} \frac{\Gamma (\lambda + 5/2 )}{\Gamma (\lambda + 4)}\,,
\end{equation}
where $\lambda$ is the same one found in Eq.~\eqref{realpart}.

%-------------------------------------
\subsection{Partial dipole amplitude}
%-------------------------------------

The partial dipole amplitude
is often used as a universal ingredient of the $\gamma p$ amplitude, as suggested by Refs. \cite{Kowalski:2003hm,Rezaeian:2013tka, Kowalski:2006hc,Rezaeian:2012ji,Kopeliovich:2001xj,Kopeliovich:1993pw,Mantysaari:2017dwh,Mantysaari:2016jaz,Goncalves:2004bp,Cepila:2019skb} . With the purpose of scanning the impact-parameter profile of the proton target we tested five different models for the partial dipole amplitude available in the literature and then selected the two of them that best describe the available data from the HERA collider, namely, the impact parameter ($b$) dependent dipole saturation model \cite{Kowalski:2003hm} (or bsat) and the $b$-dependent color glass condensate model \cite{Rezaeian:2013tka} (known as bCGC). 

In the framework of bsat model, we utilize the following formula for the dipole amplitude,
\begin{equation}
    N (x, \mathbf{r}, \mathbf{b}) = 1 - \exp \left( -\frac{\pi^2}{2 N_c}\, r^2\, \alpha_s(\mu^2)\, xg(x, \mu^2)\, T(b)\right)\,,
\end{equation}
where $\mu^2 = \mu_0^2 + 4/r^2 $ is the factorization scale in the gluon parton distribution function (PDF). In our numerical calculations, we have used the CT14LO parameterization \cite{Dulat:2015mca}, motivated by our earlier analysis of photoproduction cross sections performed in Refs.~\cite{Henkels:2020kju, Henkels:2020qvo}. This approach considers a different gluon PDF from the original fit of the bsat model (which is a gluon evolution without considering its coupling to quarks), but the numerical results are similar enough to neglect the differences. Besides, the original fit of the bsat gluon distribution does not take into account a correction to the phase factor shown in Eq.~\eqref{hatta correction}. Here, we consider the conventional Gaussian form for the proton profile function,
\begin{equation}
    T(b) = \frac{1}{2 \pi B_{\rm G}} e^{-b^2/ 2 B_{\rm G}}\,,
\end{equation}
where the parameter $B_{\rm G} = 4.25 \, \mathrm{GeV}^{-2}$ is found in Ref.~\cite{Kowalski:2003hm}.

The bCGC model interpolates two well known evolution equations: the Balitsky-Fadin-Kuraev-Lipatov (BFKL) equation near the saturation regime and Balitsky-Kovchegov (BK) equation for the saturated case. This model predicts the partial dipole amplitude in the following form:
\begin{equation}
  N (x, \mathbf{r}, \mathbf{b}) = 
  \begin{cases}  N_0 (\frac{r\, Q_s}{2})^{2 [\gamma_s + (1/(\eta \Lambda Y)) \ln (2/r Q_s)]} \,, & rQ_s \leq 2  \\
  1 - e^{-A \ln^2(B\, r\, Q_s)} \,, & rQ_s > 2
  \end{cases} \,,
\end{equation}
where $Y = \ln (1/x)$, and 
\begin{equation}
    Q_s \equiv Q_s (x, b) = \left( \frac{x_0}{x}\right)^{\Lambda/2} \left[ \exp \left( -\frac{b^2}{2 B_{\rm CGC}} \right)  \right]^{1/(2 \gamma_s)}\,,
\end{equation}
is the saturation scale which depends on the impact parameter $b$. The coefficients $A$ and $B$ are determined by the continuity condition and the other free parameters $N_0,\gamma_s,\eta,x_0,\lambda,B_{\rm CGC}$ were fixed by fitting the HERA data. In this work, we have used the corresponding parameterisation from Ref.~\cite{Rezaeian:2013tka}.

%----------------------------------------------------
\subsection{Holographic vector meson wave functions}
%----------------------------------------------------

Following Ref.~\cite{Forshaw:2012im}, we employ a semiclassical approximation to light-front QCD, where the vector meson wave function can be written in the factorized form,
\begin{equation}
    \phi (\beta, \zeta, \varphi) = \frac{\Phi(\zeta)}{\sqrt{2 \pi \zeta}} f(\beta) e^{iL\varphi} \, ,
    \label{fact-rho}
\end{equation}
where $L$ is the orbital quantum number and $\zeta = \sqrt{\beta(1-\beta)} r$. The function $\Phi(\zeta)$ satisfies the relativistic Schr\"odinger equation,
\begin{equation}
    \left( -\frac{d^2}{d \zeta^2} - \frac{1 - 4L^2}{4 \zeta^2} + U(\zeta)\right) \Phi(\zeta) = M^2 \Phi(\zeta) \,,
    \label{schro-rho}
\end{equation}
where $U(\zeta)$ is the confining potential defined in the light-front frame. Here, we employ the soft-wall model,
\begin{equation}
    U (\zeta) = \kappa^4 \zeta^2 + 2 \kappa^2(J-1) \,.
\end{equation}
The eigenvalues of the Schr\"odinger equation are
\begin{equation}
    M^2 = 4 \kappa^2 \left(n + \frac{J}{2} + \frac{L}{2}\right)\,
    \label{eq:mass_eigenvalue}
\end{equation}
and, in order to fix $\kappa$, the eigenvalue with $n=0$, $J=1$, and $L=0$ is compared to the ground state vector meson mass squared, i.e., $\kappa = M_V/\sqrt{2}$.

In the original wave function model reviewed in \cite{Brodsky:2014yha}, $\kappa$ is a universal constant parameter which does not change from one light vector meson to other. As a consequence, in order to describe the spectroscopy measurements, it is necessary to introduce a mass shift in Eq.~(\ref{eq:mass_eigenvalue}). This provides good results for the ground state mesons with $L=0$ and $L=2$, but excited states with $L=0$ have results not so precise. These $L=0$ excited states are the ones we are interested here. Thus we found that using different values of $\kappa$ depending on each meson family ($M_\rho$, $M_\omega$ and $M_\phi$ from recent Ref.~\cite{ParticleDataGroup:2016lqr}), we get a good description of the $L=0$ spectroscopy. The evidence of nonuniversal $\kappa$ in the case of heavy mesons is discussed in \cite{Brodsky:2014yha}. Also the $\kappa$ parameter for pions is different, as it can be seen e.g. in Ref.~\cite{Vega:2009zb}. Even for light vector mesons different values of $\kappa$ can be extrated from the Regge slope in a similar fashion as in Ref.~\cite{Lee:2018zud}.

The specification of the function $f(\beta)$ in Eq.~\eqref{fact-rho} is done by comparing the expressions to the pion electromagnetic (EM) form factor, as can be seen in Ref.~\cite{Brodsky:2007hb}, it takes the form $f(\beta) \sim \sqrt{\beta (1-\beta)}$. Solving Eq.~\eqref{schro-rho} we get the dynamical part of the AdS/QCD wave function, which gets us to
\begin{equation}
   \Phi_{n,L} (\zeta) =  \kappa^{1+L} \sqrt{\frac{2n!}{(n+L)!}} \zeta^{1/2 + L} \exp \left( \frac{-\kappa^2 \zeta^2}{2}\right) L_n^L (\kappa^2 \zeta^2)\,.
\end{equation}

In order to incorporate the quark mass, we employ the Brodsky-Téramond treatment \cite{Brodsky:2014yha} that extends the transverse momentum dependence in order to include the full invariant mass. For massive quarks, the basic working ansatz is to replace $M$ by the $q\bar{q}$ invariant mass, $M_{q\bar{q}}$, as
\begin{equation}
    M^2 = \frac{\textbf{k}_{\perp}^2}{\beta(1-\beta)} \rightarrow M^2_{q \bar{q}} = \frac{\textbf{k}^2_{\perp}}{\beta(1-\beta)} + \frac{m_q^2}{\beta} + \frac{ m^2_{\bar q}}{1-\beta}\,,
\end{equation}
where $q (\bar{q})$ is the quark (antiquark) flavor, and $\textbf{k}_{\perp}$ is the relative quark transverse momentum. Thus, the dynamical part of the resulting light-front wave function (LFWF) is modified by the inclusion of an exponential factor for the quark masses, as follows
\begin{equation}
    \phi_{n, L} (\beta, \zeta )\sim \sqrt{\beta(1-\beta)} e^{\frac{1}{2 \kappa^2} \left(\frac{m_q^2}{\beta} + \frac{m^2_q}{1-\beta}\right)} \zeta^2 e^{-\frac{1}{2}\kappa^2 \zeta^2} L_n^L (\kappa^2 \zeta^2)\,,
    \label{eq:scalar_part_wf}
\end{equation}
where $L_n^L (\kappa^2 \zeta^2)$ are the Laguerre polynomials.

There is still a large uncertainty regarding the effective quark masses, which also incorporates non-perturbative effects. Following Refs.~\cite{Forshaw:2011yj}, we considered $m_{u,d} = 0.14 $ GeV as the universal mass scale for the up and down quarks, while for the strange quark we considered a larger value $m_s = 0.35 $ GeV, which is close to values used in other analyses \cite{Brodsky:2014yha, Cisek:2010jk, Bolognino:2019pba}. It is good to mention that we tested some different values for the quark masses and we found that there is a modest dependence of the results on these parameter values. For this reason, we chose to use the light quark mass values obtained from the fitting of the dipole models bCGC and bSat to mainly HERA data. 

Since the previous equations only describe the scalar part of the wave functions, now we take into account the helicities of the quark ($h$) and antiquark ($\bar h$). Following Refs.~\cite{Kaur:2020emh,Kowalski:2006hc}, one can express the complete wave function as follows 
\begin{equation}
{\Psi_{V, L}^{h, \bar{h}}}(r, \beta)= \mathcal{N}_L \frac{1}{2 \sqrt{2}} \delta_{h,-\bar{h}}\left(1+\frac{m_f^2-\nabla^2_r}{M^2 \beta(1-\beta)}\right) \phi(\beta, \zeta)
\end{equation}
for the longitudinal part, and
\begin{equation}
{\Psi_{V, T= \pm}^{h, \bar{h}}}(r, \beta)= \pm \mathcal{N}_T \left[i e^{ \pm i \theta}\left(\beta \delta_{h \pm, \bar{h} \mp}-(1-\beta) \delta_{h \mp, \bar{h} \pm}\right) \partial_r+m_f \delta_{h \pm, \bar{h} \pm}\right] \frac{\phi(\beta, \zeta)}{2 \beta(1-\beta)} \, .
\end{equation}
for the transversal part, respectively, where $\phi (\beta, \zeta)$ is the scalar part of the wave function given by Eq.~(\ref{eq:scalar_part_wf}). It is important to emphasize that these are frame-independent light-front wave functions normalized to unity, the eigensolutions of the LF Hamiltonian defined at fixed LF time.

%--------------------------
\section{Numerical results}
\label{sec:results}
%--------------------------

The formalism presented above enables us to perform an analysis of total and differential cross sections, as will be shown hereafter, for the three light vector mesons $\rho$, $\omega$ and $\phi$ production in $\gamma^{(*)} p$ collisions. As was mentioned earlier, we use the AdS/QCD holographic wave functions in all numerical results presented below. The hadron-level cross sections are very sensitive to the details of modelling of the color dipole interaction with the proton target, and hence to the corresponding parametrization of the partial dipole amplitude \cite{Henkels:2020qvo, Goncalves:2022wzq}. Such a large sensitivity arises mostly from dominant soft and non-perturbative kinematic domains poorly constrained by traditional fits of the dipole parametrizations to the hard DIS data from HERA. In the current work, we choose to show numerical results obtained with the bCGC and bsat dipole models, which in our analysis provide the best description of the available data on ground-state $\rho$, $\omega$ and $\phi$ photoproduction cross sections.
%---------------------
\begin{figure}[bt]
\begin{minipage}{0.48\textwidth}
 \centerline{\includegraphics[width=1.0\textwidth]{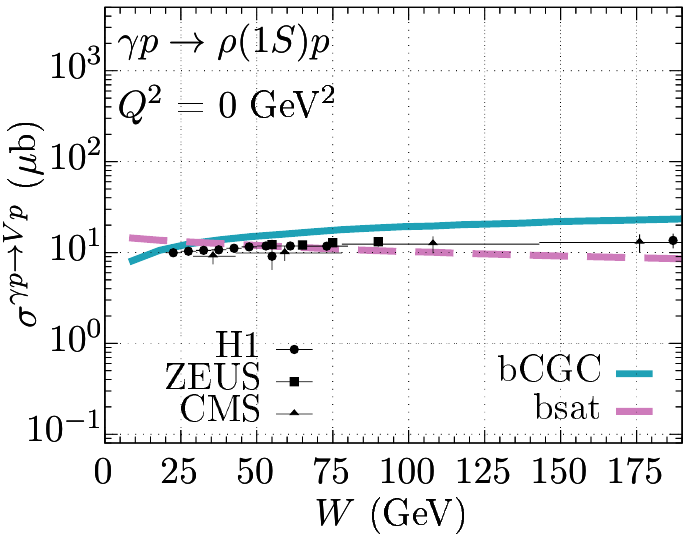}}
\end{minipage} \hfill
\caption{Total cross section for the $\rho(1S)$ photoproduction as a function of $\gamma p$ center-of-mass energy $W$ obtained by using the holographic wave function with the bCGC and bsat dipole models. The results are compared to the corresponding data by H1~\cite{H1:1996prv,H1:2020lzc}, ZEUS~\cite{ZEUS:1997rof} and CMS~\cite{CMS:2019awk} collaborations. }
\label{fig:rho-photoproduction}
\end{figure}
%---------------------

Fig.~\ref{fig:rho-photoproduction} shows the total cross section of $\rho$-meson photoproduction as a function of the $\gamma p$ center-of-mass energy $W$. The results were obtained using the holographic wave function with the bCGC and bsat $b$-dependent dipole parametrizations. In this figure, were also included the experimental data from H1~\cite{H1:1996prv,H1:2020lzc}, ZEUS~\cite{ZEUS:1997rof} and CMS~\cite{CMS:2019awk} collaborations. Apparently, the bsat model provides a better description of the available data in comparison to the results obtained with the bCGC model, particularly, at smaller $W$ values. For higher $W$, however, the bsat model somewhat underestimates the data points while the bCGC model overestimates them. 
%---------------------
\begin{figure}[tb]
\begin{minipage}{0.48\textwidth}
 \centerline{\includegraphics[width=1.0\textwidth]{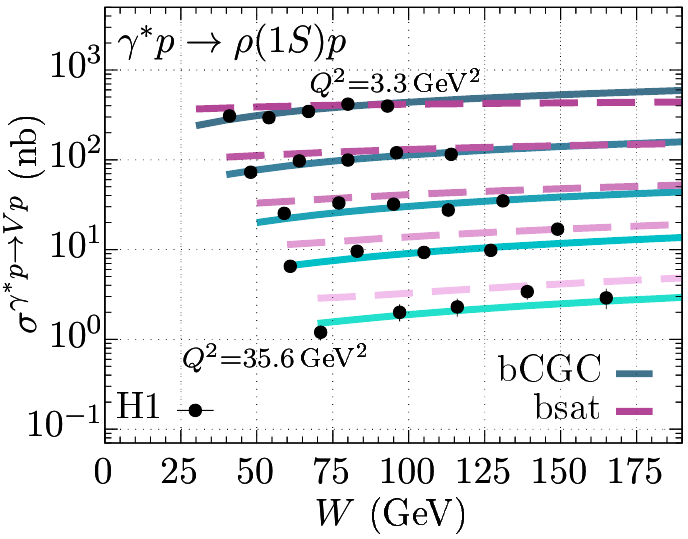}}
\end{minipage} \hfill
\begin{minipage}{0.48\textwidth}
 \centerline{\includegraphics[width=1.0\textwidth]{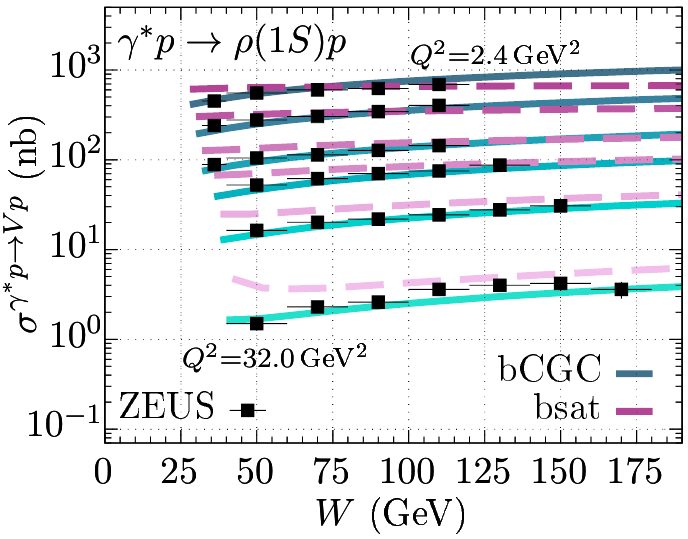}}
\end{minipage} \hfill
\caption{Total cross section for $\rho (1S)$ electroproduction as a function of $W$ obtained by using the holographic wave function, together with the bCGC and bsat dipole models. The results are compared, on the left, to the data from the H1~\cite{H1:2009cml} collaboration shown for five values of $Q^2$ (from top to bottom, we have $Q^2 = $ $3.3$, $6.6$, $11.9$, $19.5$ and $35.6$ GeV\textsuperscript{2}, respectively) and, on the right, to the data from ZEUS~\cite{ZEUS:2007iet} collaboration for six different values of $Q^2$ (from top to bottom, $Q^2 = $ $2.4$, $3.7$, $6.0$, $8.3$, $13.5$ and $32.0$ GeV\textsuperscript{2}, respectively).}
\label{fig:rho-electroproduction}
\end{figure}
%---------------------

Fig.~\ref{fig:rho-electroproduction} also presents the total cross section as a function of $W$, however, in variance to Fig.~\ref{fig:rho-photoproduction}, it is calculated for $\rho$ electroproduction with non-zeroth photon virtualities $Q^2$. Here, the darker curves are given for small $Q^2$ values while the lighter ones correspond to higher $Q^2$. Again, the results were obtained by using the holographic wave function, as well as with the bCGC and bsat models, and compared, on the left panel, to the H1 data \cite{H1:2009cml} for five distinct values of $Q^2$ (from top to bottom, $Q^2 = $ $3.3$, $6.6$ , $11.9$, $19.5$ and $35.6$ GeV\textsuperscript{2}, respectively) and, on the right panel, to the ZEUS data \cite{ZEUS:2007iet} for six values of $Q^2$ (from top to bottom, $Q^2 = $ $2.4$, $3.7$, $6.0$, $8.3$, $13.5$ and $32.0$ GeV\textsuperscript{2}, respectively). We notice that the bCGC model appears to be the most successful in description of the experimental data for all available values of $Q^2$ and in all measured $W$ ranges.
%---------------------
\begin{figure}[tb]
\begin{minipage}{0.48\textwidth}
 \centerline{\includegraphics[width=1.0\textwidth]{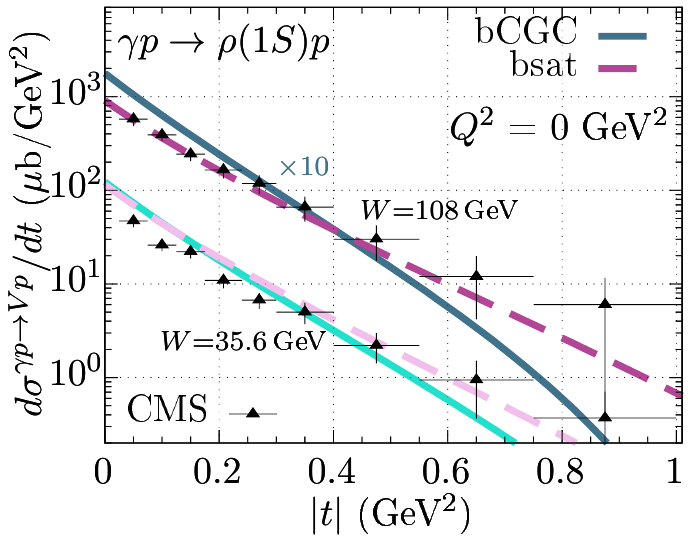}}
\end{minipage} \hfill
\begin{minipage}{0.48\textwidth}
 \centerline{\includegraphics[width=1.0\textwidth]{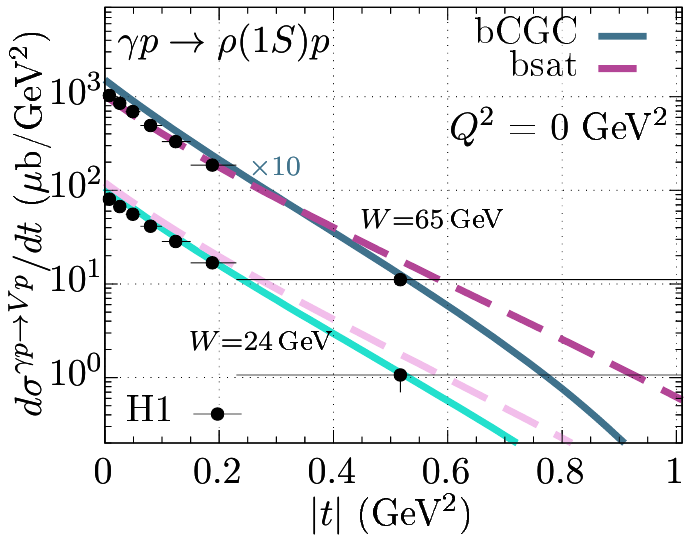}}
\end{minipage} \hfill
\caption{Differential cross section of $\rho(1S)$ photoproduction as a function of the momentum transfer squared $|t|$ obtained with the bCGC and bsat dipole models for different values of $W$ and compared to the corresponding data from the CMS collaboration \cite{CMS:2019awk} (left panel) and to those from the H1 collaboration \cite{H1:2020lzc} (right panel). The darker upper curves (for $W = 108$ GeV) were multiplied by a factor of ten in order to distinguish them from the lighter ones (for $W = 65$ GeV).}
\label{fig:rho-dt-photoproduction}
\end{figure}
%---------------------

Besides the total cross sections, we also make use of the dipole formalism to estimate the corresponding differential cross sections. It is worth mentioning that b-dependent parameterizations are crucial for obtaining these observables. In Fig.~\ref{fig:rho-dt-photoproduction} we show the differential cross section for the $\rho(1S)$ photoproduction ($Q^2 = 0$ GeV$^2$) as a function of the momentum transfer squared $|t|$ for $W = 35.6, 108$ GeV (left panel), and for $W = 24, 65$ GeV (right panel), in comparison to the corresponding data from CMS \cite{CMS:2019awk} and H1 \cite{H1:2020lzc} collaborations, respectively. Again, here we apply the holographic wave functions and the bCGC and bsat dipole parametrizations. In order to avoid an overlap of the curves, in each panel, we multiplied the curves with higher $W$ values ($W = 108$ GeV, $W = 65$ GeV), represented by darker colors, by a factor of ten. One may notice that the bsat model provides a better overall description of all the available data sets for higher $|t|$ values. On the other hand, one should note also that the largest-$|t|$ data points from the H1 Collaboration have big uncertainties and that the bCGC model comes closer to the central values of the measurement. At small $|t|$, one can see that the bsat model features a better description of the data for higher $W$ (darker curves). Interestingly enough, the curve for the bCGC model comes very close to the $W = 24$ GeV data points, while the curves for both bCGC and bsat models pass through the $W = 35.6$ GeV data points. This observation emphasizes the fact that there is not a single $b$-dependent partial dipole amplitude parametrisation that perfectly describes all existing $\rho(1S)$ photoproduction measurements.
%---------------------
\begin{figure}[tb]
\begin{minipage}{0.48\textwidth}
 \centerline{\includegraphics[width=1.0\textwidth]{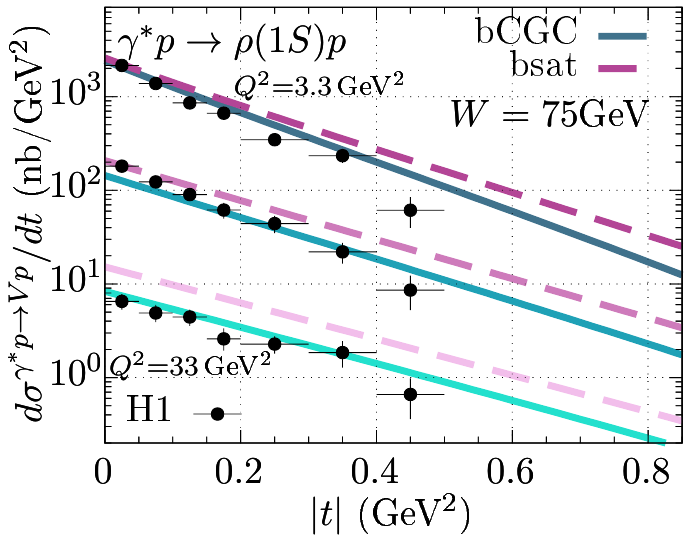}}
\end{minipage} \hfill
\begin{minipage}{0.48\textwidth}
 \centerline{\includegraphics[width=1.0\textwidth]{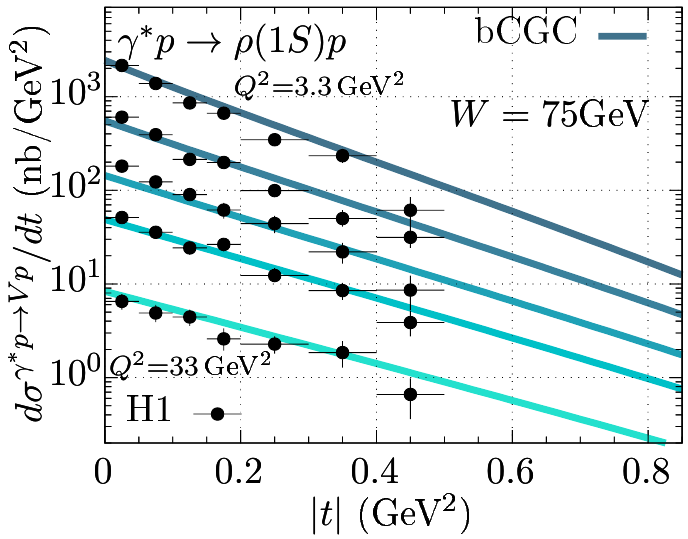}}
\end{minipage} \hfill
\caption{Differential cross section for $\rho(1S)$ meson electroproduction as a function of the momentum transfer squared $|t|$ for $W = 75$ GeV. The left panel shows a comparison of the results obtained by using the bCGC and bsat models with the H1 data \cite{H1:2009cml} for three distinct values of $Q^2$ (from top to bottom, $Q^2 = $ $3.3$, $11.5$ and $33.0$ GeV\textsuperscript{2}, respectively). The right panel presents the curves obtained only with the bCGC model and compared to the H1 data for five different $Q^2$ values (from top to bottom, $Q^2 = $ $3.3$, $6.6$, $11.5$, $17.4$ and $33.0$ GeV\textsuperscript{2}, respectively).}
\label{fig:rho-dt-electroproduction}
\end{figure}
%---------------------

In Fig.~\ref{fig:rho-dt-electroproduction} we present the differential cross section as a function of the momentum transfer squared $|t|$ for $\rho(1S)$ electroproduction at $W = 75$ GeV. On the left panel, a comparison between the results obtained using the bCGC and bsat models and the corresponding H1 data \cite{H1:2009cml} for three different values of $Q^2$ (from top to bottom, $Q^2 = $ $3.3$, $11.5$ and $33.0$ GeV\textsuperscript{2}, respectively). One notices here that the bCGC model performs somewhat better in describing the experimental data. Thus, on the right panel, we show the results obtained only with the bCGC model, but compared with all five available datasets at five distinct $Q^2$ values (namely, from top to bottom, $Q^2 = $ $3.3$, $6.6$, $11.5$, $17.4$ and $33.0$ GeV\textsuperscript{2}, respectively).

\begin{figure}[tb]
\begin{minipage}{0.48\textwidth}
 \centerline{\includegraphics[width=1.0\textwidth]{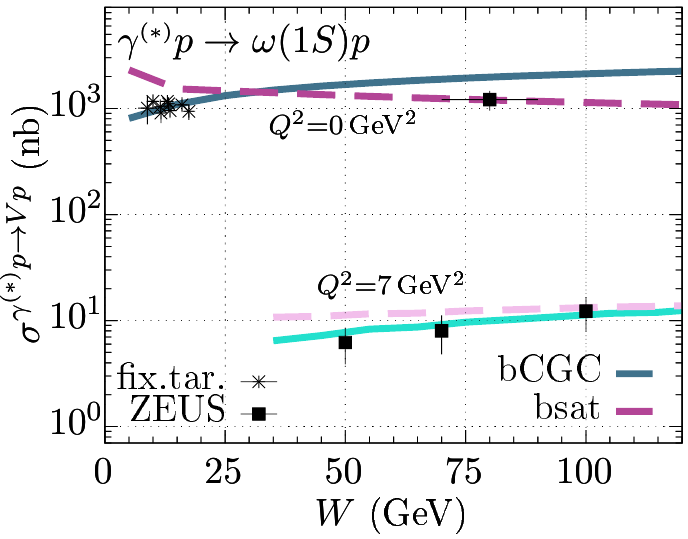}}
\end{minipage} \hfill
\begin{minipage}{0.48\textwidth}
 \centerline{\includegraphics[width=1.0\textwidth]{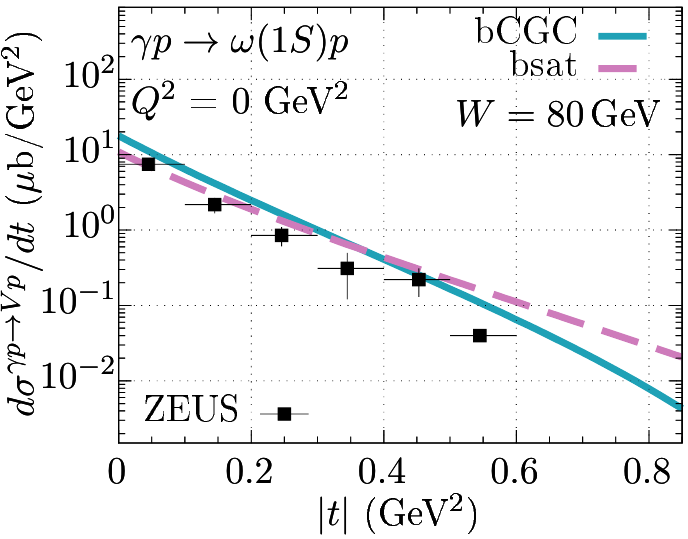}}
\end{minipage} \hfill
   \caption{Results for the $\omega(1S)$ photo- and electroproduction cross sections obtained with the use of the holographic wave function, as well as with the bCGC and bsat dipole models. On the left panel, the total cross section is shown as a function of $W$ for $Q^2 = 0$ GeV\textsuperscript{2} (darker curve) and $Q^2 = 7$ GeV\textsuperscript{2} (lighter curve) in comparison with the fixed target measurements \cite{BrownHarvardMITPadovaWeizmannInstituteBubbleChamberGroup:1967zz,Aachen-Berlin-Bonn-Hamburg-Heidelberg-Munich:1968rzt,Davier:1969nx,Ballam:1971wq,Ballam:1972eq,Aachen-Hamburg-Heidelberg-Munich:1975jed,Egloff:1979xg,Breakstone:1981wk,Bonn-CERN-EcolePoly-Glasgow-Lancaster-Manchester-Orsay-Paris-Rutherford-Sheffield:1982qiv,LAMP2Group:1984qvm,OmegaPhoton:1983huz,Busenitz:1989gq} (a compilation of these data can be found in Ref.~\cite{ZEUS:1996zse}) and also with the data from the ZEUS Collaboration~\cite{ZEUS:1996zse,ZEUS:2000swq}. On the right panel, the differential cross section is shown as a function of momentum transfer squared $|t|$ for $W = 80$ GeV in comparison to the ZEUS data~\cite{ZEUS:1996zse}.}
   \label{fig:omega-photo-electroproduction}
\end{figure}

\begin{figure}[tb]
\begin{minipage}{0.48\textwidth}
 \centerline{\includegraphics[width=1.0\textwidth]{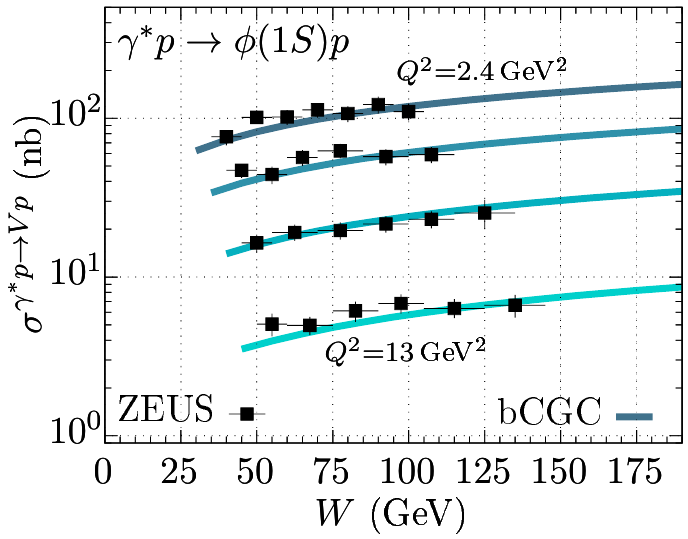}}
\end{minipage} \hfill
\begin{minipage}{0.48\textwidth}
 \centerline{\includegraphics[width=1.0\textwidth]{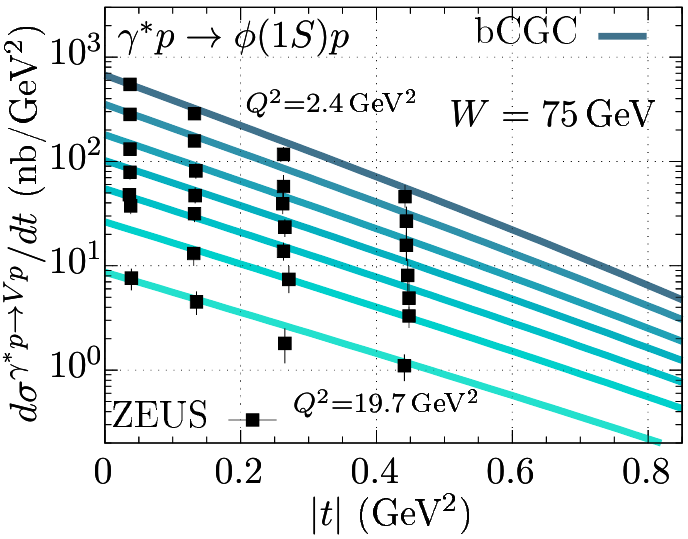}}
\end{minipage} \hfill
  \caption{Results for the $\phi(1S)$ electroproduction cross sections compared with the ZEUS data~\cite{ZEUS:2005bhf}. Here, we show the results only for the bCGC dipole parametrization. On the left panel, the total cross section is shown as a function of $W$ in comparisons to the four datasets with different $Q^2$ values (from top to bottom, $Q^2 = $ $2.4$, $3.8$, $6.5$ and $13.0$ GeV\textsuperscript{2}, respectively). On the right panel, the differential cross section is shown as a function of $|t|$ for $W = 75$ GeV versus data points for seven different values of $Q^2$ (from top to bottom, $Q^2 = $ $2.4$, $3.6$, $5.2$, $6.9$, $9.2$, $12.6$ and $19.7$ GeV\textsuperscript{2}, respectively).}
  \label{fig:phi-electroproduction}
\end{figure}

As was mentioned earlier, our work aims at exploiting the holographic approach for light vector meson wave functions which can be used to obtain the observables for other mesons than $\rho(1S)$. In Figs.~\ref{fig:omega-photo-electroproduction} and \ref{fig:phi-electroproduction}, we present the cross sections for $\omega(1S)$ and $\phi(1S)$ production, respectively. 

On the left panel of Fig.~\ref{fig:omega-photo-electroproduction} we display the total cross section as a function of $W$ for two different cases: $Q^2 = 0$ GeV\textsuperscript{2} (darker curve) and for $Q^2 = 7$ GeV\textsuperscript{2} (lighter curve), in comparison to the fixed target data~\cite{BrownHarvardMITPadovaWeizmannInstituteBubbleChamberGroup:1967zz,Aachen-Berlin-Bonn-Hamburg-Heidelberg-Munich:1968rzt,Davier:1969nx,Ballam:1971wq,Ballam:1972eq,Aachen-Hamburg-Heidelberg-Munich:1975jed,Egloff:1979xg,Breakstone:1981wk,Bonn-CERN-EcolePoly-Glasgow-Lancaster-Manchester-Orsay-Paris-Rutherford-Sheffield:1982qiv,LAMP2Group:1984qvm,OmegaPhoton:1983huz,Busenitz:1989gq} (a compilation of these data can be found in Ref.~\cite{ZEUS:1996zse}), as well as to the data from the ZEUS Collaboration \cite{ZEUS:1996zse,ZEUS:2000swq}. As can be seen, the bCGC model describes all available electroproduction data rather well (i.e. excluding the ZEUS data point for $Q^2 = 0$ GeV\textsuperscript{2}). 

On the right panel of Fig.~\ref{fig:omega-photo-electroproduction}, the differential cross section for the $\omega(1S)$ photoproduction is shown as a function of momentum transfer squared $|t|$ for $W = 80$ GeV in comparison to the data from the ZEUS Collaboration~\cite{ZEUS:1996zse}. The curve comes very close to and features a similar shape as the experimental data. This is a rather important observation given a practical challenge in description of all $t$-dependent differential cross sections in the framework of a single dipole parametrization. 
%---------------------
%---------------------

On the left panel of Fig.~\ref{fig:phi-electroproduction}, the total cross section $\phi(1S)$ electroproduction is presented as a function of $W$ in comparison with the experimental data from the ZEUS Collaboration~\cite{ZEUS:2005bhf}. Here, we show the results only for the bCGC dipole parametrization, the most successful one in description of the vector meson electroproduction data. As was the case for other vector mesons, the curves describe the four data sets available for different $Q^2$ values rather well (from top to bottom, $Q^2 = $ $2.4$, $3.8$, $6.5$ and $13.0$ GeV\textsuperscript{2}, respectively). On the right panel, the differential cross section is shown as a function of $|t|$ for $W = 75$ GeV. Likewise, the data for all seven available data sets for different $Q^2$ values provided by the ZEUS Collaboration~\cite{ZEUS:2005bhf} are described pretty well (from top to bottom, $Q^2 = $ $2.4$, $3.6$, $5.2$, $6.9$, $9.2$, $12.6$ and $19.7$ GeV\textsuperscript{2}, respectively). It is worth mentioning that using a vector meson mass dependent $\kappa$ parameter made it possible to not only describe all the available $\rho(1S)$ and $\omega(1S)$ data points but also to describe well the existing measurements of $\phi(1S)$. So one may conclude here that the considered mass dependence of $\kappa$ in the effective confining potential provides a good description of the experimental data sets for all three light vector mesons. 
%---------------------
%---------------------

\begin{figure}[tb]
\begin{minipage}{0.48\textwidth}
 \centerline{\includegraphics[width=1.0\textwidth]{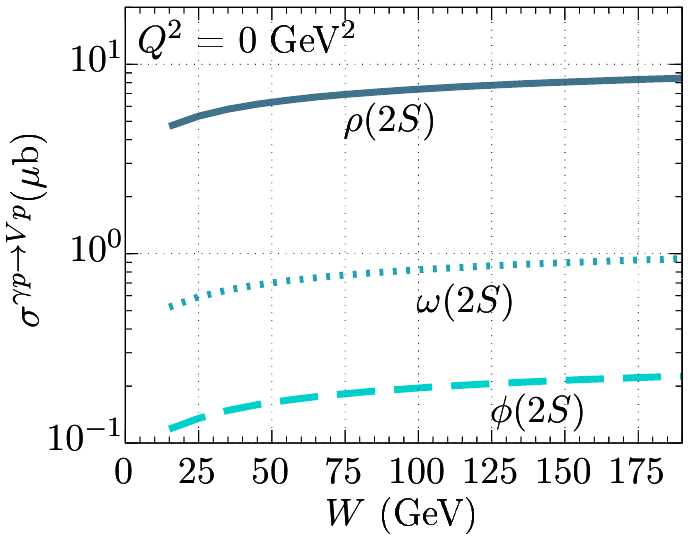}}
\end{minipage} \hfill
\begin{minipage}{0.48\textwidth}
 \centerline{\includegraphics[width=1.0\textwidth]{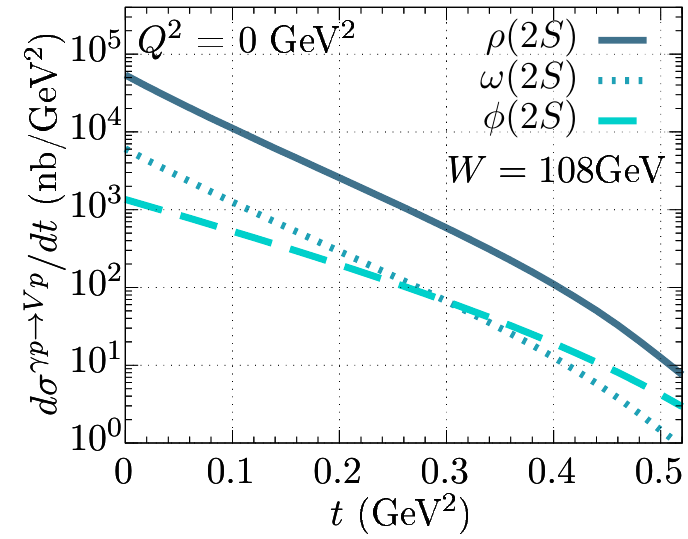}}
\end{minipage} \hfill
  \caption{Predictions for the total photoproduction cross section as a function of $W$ (left panel) and for the differential cross section as a function of momentum transfer squared $|t|$ for $\rho(2S)$ (darker blue solid line), $\omega(2S)$ (medium shade of blue dotted line) and $\phi(2S)$ (lighter blue dashed line) mesons.}
  \label{fig:excited_states_photoproduction}
\end{figure}

\begin{figure}[tb]
\begin{minipage}{0.48\textwidth}
 \centerline{\includegraphics[width=1.0\textwidth]{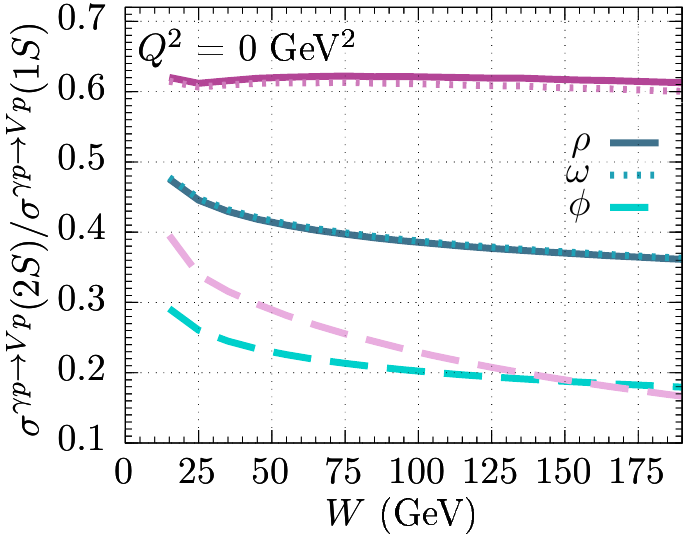}}
\end{minipage} \hfill
\begin{minipage}{0.48\textwidth}
 \centerline{\includegraphics[width=1.0\textwidth]{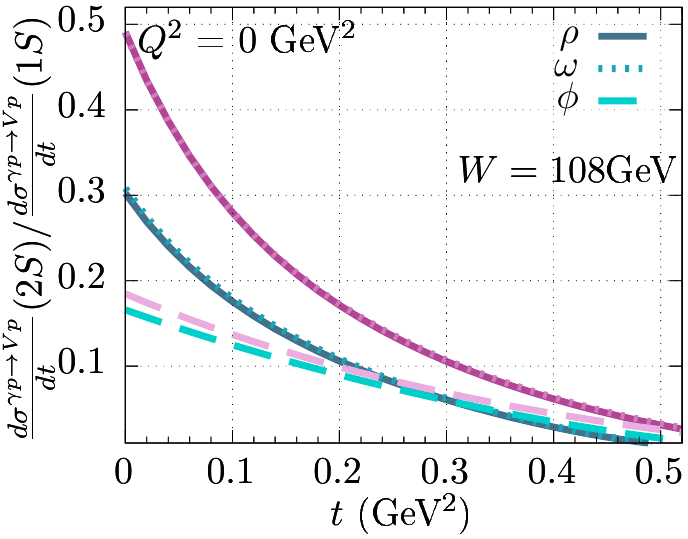}}
\end{minipage} \hfill
  \caption{
  Predictions for the ratio of the excited-state total cross section to the ground-state total cross section as a function of $W$ (left panel) and for that of the excited-state differential cross section to the ground-state differential cross section as a function of momentum transfer squared $|t|$ (right panel) for $\rho$ (solid lines), $\omega$ (dotted lines) and $\phi$ (dashed lines) mesons. The blue curves are obtained with the bCGC model, while the violet ones correspond to the bsat model. 
  }
  \label{fig:excited_states_ratio-photoproduction}
\end{figure}

Finally, the holographic wave functions approach enables us to make predictions for the photo- and electroproduction cross sections for various vector meson excited states. Here, we present on the left panel of Fig.~\ref{fig:excited_states_photoproduction} the predictions for the total photoproduction cross section for $\rho(2S)$ (darker blue solid line), $\omega(2S)$ (medium shade of blue dotted line) and $\phi(2S)$ (lighter blue dashed line) mesons as functions of $W$. On the right panel, we show the corresponding predictions for the differential cross sections as functions of $|t|$ for a fixed $W = 108$ GeV. All these curves are obtained with the bCGC model -- the most successful in description of the ground-state electroproduction data (see above). Since there are large discrepancies between the results obtained with different parametrizations for the partial dipole amplitude, mainly for photoproduction processes, we chose to show in Fig.~\ref{fig:excited_states_ratio-photoproduction} the predictions for the ratio of the excited-state total cross section to the corresponding ground-state total cross section as a function of $W$ (left panel) as well as the ration of the excited-state differential cross section to the ground-state differential cross section as a function of $|t|$ for $W = 108$ GeV (right panel) for the three different light vector mesons. In order to make the visualization of the curves clearer, we use darker solid lines for $\rho$, medium shade dotted lines for $\omega$ and lighter dashed lines for $\phi$. Also, we utilize blue shades to represent the curves obtained with the use of the bCGC model and violet shades for the ones obtained with the bsat model. As can be seen on both panels, the results obtained with the bsat model are much higher than the ones obtained with the bCGC parametrization. This result illustrates the statement that there is still big uncertainties in the structure of partial dipole amplitude, primarily, in the soft and nonperturbative domain \cite{SampaiodosSantos:2014qtt} and that some new improved parametrizations are required in order to describe all exclusive processes for light vector meson production. The future measurements of excited states' photoproduction could play a significant role in further constraining the dipole model in the nonperturbative range.
%---------------------
%---------------------
%---------------------
%--------------------------

%--------------------------
\section{Conclusion}
\label{sec:conclusions}
%--------------------------
    In this work, the exclusive photo- and electroproduction of the light vector ($\rho$, $\omega$, $\phi$) mesons are studied within the color dipole picture. By using the bCGC dipole amplitude it was possible to obtain a very good description of the available data for the electroproduction cross sections of all three light mesons in the ground state $\rho(1S)$, $\omega(1S)$, and $\phi(1S)$. For the nonperturbative meson wave function, the light front holographic QCD model was used, where the wave function is the solution of a relativistic equation that coincides with the Schroedinger equation with a confining potential. It proved to be important for the description of the $\phi$ cross section as well as its spectroscopy to have a vector meson mass-dependent $\kappa$ parameter in the effective confining potential. 
    
    For the photoproduction case, we calculated the differential cross section, with the same setup, and obtained a good description of the available ZEUS data for the $\omega(1S)$ production and for the CMS data for $\rho(1S)$ production at small $t$. In the case of the $\rho(1S)$ photoproduction total cross section, we showed that there is a large sensitivity to the $b$-dependent dipole amplitude model employed, highlighting that the construction of such models could benefit from this data and also from future measurements.
    
    The light-front holographic QCD model provides not only the light vector mesons ground-state wave function, but also that for the excited states. We have evaluated the cross section of photoproduction of the excited states, i.e, $\rho(2S)$, $\omega(2S)$, and $\phi(2S)$. Again, there were differences between the predictions obtained with the two partial dipole amplitude models. Measurements of these observables, such as the ones coming from the future FoCal detector \cite{Bylinkin:2022wkm}, could enhance our understanding about this nonperturbative part of the color dipole approach and help us to improve its determination. 

%--------------------------
\section*{Acknowledgments}
%--------------------------

This work was supported by Fapesc, INCT-FNA (464898/2014-5), and CNPq (Brazil) for CH, EGdO, and HT. 
This study was financed in part by the Coordenação de Aperfeiçoamento de Pessoal de Nível 
Superior -- Brasil (CAPES) -- Finance Code 001. The work has been performed in 
the framework of COST Action CA15213 ``Theory of hot matter and relativistic heavy-ion collisions''
(THOR). R.P.~is supported in part by the Swedish Research Council grants, contract numbers
621-2013-4287 and 2016-05996, as well as by the European Research Council (ERC) under 
the European Union's Horizon 2020 research and innovation programme (grant agreement No 668679). 

\bibliography{bib}

\end{document}